\documentclass[12pt]{article}

\usepackage{amsbsy,amssymb,latexsym,amsfonts,amsmath}
\usepackage{mathrsfs}
\usepackage{graphicx}
\usepackage{bm}
\usepackage{here}

\numberwithin{equation}{section}
\newcommand{\bel}[1]{\begin{equation}\label{#1}}                     
\newcommand{\bal}[1]{\begin{eqnarray}\label{#1}}                     
\newcommand{\be}{\begin{equation}}
\newcommand{\ee}{\end{equation}}

\newcommand{\im}{\mathrm{i}}
\newcommand{\ex}{\mathrm{e}}
\newcommand{\de}{\mathrm{d}}

\newcommand{\qq}{\qquad}

\renewcommand{\thefootnote}{\fnsymbol{footnote}}

\newcommand{\bea}{\begin{equation}}
\newcommand{\eea}{\end{equation}}

\begin{document}
%
%
\begin{titlepage}
\begin{flushright}
\normalsize
~~~~
OCU-PHYS 405\\
August, 2014 \\
\end{flushright}

\vspace{15pt}

\begin{center}
{\LARGE $q$-Virasoro/W Algebra at Root of Unity and Parafermions}
\end{center}

\vspace{23pt}

\begin{center}
{ H. Itoyama$^{a, b}$\footnote{e-mail: itoyama@sci.osaka-cu.ac.jp},
T. Oota$^b$\footnote{e-mail: toota@sci.osaka-cu.ac.jp}
  and  R. Yoshioka$^b$\footnote{e-mail: yoshioka@sci.osaka-cu.ac.jp}  }\\
%
\vspace{18pt}
%

$^a$ \it Department of Mathematics and Physics, Graduate School of Science\\
Osaka City University\\
\vspace{5pt}

$^b$ \it Osaka City University Advanced Mathematical Institute (OCAMI)

\vspace{5pt}

3-3-138, Sugimoto, Sumiyoshi-ku, Osaka, 558-8585, Japan \\

\end{center}
%
\vspace{20pt}
\begin{center}
Abstract\\
\end{center}
We demonstrate that the parafermions appear in the $r$-th root of unity limit of 
 $q$-Virasoro/$W_n$ algebra. 
The proper value of the central charge of the coset model
$ \frac{\widehat{\mathfrak{sl}}(n)_r \oplus \widehat{\mathfrak{sl}}(n)_{m-n}}
 {\widehat{\mathfrak{sl}}(n)_{m-n+r}}$
is given from the parafermion construction of the block in the limit.


\vfill

\setcounter{footnote}{0}
\renewcommand{\thefootnote}{\arabic{footnote}}

\end{titlepage}

\section{Introduction}

Ever since the AGT relation \cite{AGT, Wyl0907, MM0908} 
 (the correspondence between the correlators of 2d QFT and the 4d instanton sum)
was introduced, the both sides of the correspondence have been intensively studied 
 by a number of people. 
For example, in the 2d side, the $\beta$-deformed matrix model is used 
 in order to control the integral representation of the conformal block
  \cite{DV,IMO,MMS0911,MMS1001,IO5,MMM1003,IOYone}. 
There are also some proposals for proving the 2d-4d connection 
\cite{MMS1012,KMZ1308,MS1307,MMZ1312,MRZ1405}. 
Moreover similar correspondence has been found and examined
\cite{BF1105,NT1106,BBB1106,BMT,Wy1109,EPSS1110,I1110,AT1110,
BBFLT1111,BBT1211,ABT1306}. 
Among these, we pay our attention, in this paper, to the correspondence
 between the coset model, 
\bel{coset}
 \frac{\widehat{\mathfrak{sl}}(n)_{r} \oplus 
 \widehat{\mathfrak{sl}}(n)_{p}}{\widehat{\mathfrak{sl}}(n)_{r+p}}, 
\ee
and the ${\mathcal N}=2$ $SU(n)$ gauge theory on ${\bf R}^4/{\bf Z}_r$ 
\cite{Wy1109,AT1110}. 
Here $\widehat{\mathfrak{sl}}(n)_k$ stands for the affine Lie algebra 
 in the representation of level $k$ and 
 $r$ and $p$ will be specified in this paper. 

On the 2d CFT side, a quantum deformation ($q$-deformation)
of the Virasoro algebra \cite{SKAO} and the $W_n$ algebra \cite{FF95,AKOS95}
is known, while the 4d gauge theories can be lifted to five-dimensional
theories with the fifth direction compactified on a circle.
There exists a natural generalization to 
 the connection between the 2d theory 
 based on the $q$-deformed Virasoro/W algebra 
 and the five-dimensional $\mathcal{N}=2$ gauge theory \cite{AY}.  
For recent developments, see, for example, 
\cite{NPP1303,IOY2,Tan1309,Orl1310,BMPTY1310,NPPT1312,IMM1406}.  
In the previous paper \cite{IOY2}, we proposed a limiting procedure to 
 get the Virasoro/W block in the 2d side from that in the $q$-deformed version. 
On the other hand, we saw that the instanton partition function on
 ${\bf R}^4/{\bf Z}_r$ are generated from that on ${\bf R}^5$
 at the same limit. 
This result means if we assume the 2d-5d connection, 
 it is automatically assured 
 that the Virasoro/W blocks generated by using the limiting procedure agree
 with the instanton partition function on ${\bf R}^4/{\bf Z}_r$. 
Our limiting procedure corresponds to 
 a root of unity limit in $q$. A root of unity limit of the $q$-Virasoro algebra
was also considered in \cite{BP}. Our limit is slightly different from this and is
similar to the one used
 in order to construct the eigenfunctions of the spin Calogero-Sutherland model 
 from Macdonald polynomials in \cite{TU1997,Uglov}. 
 
In the present paper we will elaborate our limiting procedure and 
 show that 
 the ${\bf Z}_r$-parafermionic CFT which has the symmetry described by 
 (\ref{coset}) appears in the 2d side.  
We clarify also the relation between the free parameter $p$ and 
 the omega background parameters in the 4d side.

The paper is organized as follows: 
In the next section, we review the limiting procedure for 
 $q$-Virasoro algebra \cite{IOY2}. 
In section 3, we consider the $q$-deformed screening current and charge 
 and show that the ${\bf Z}_r$-parafermion currents are derived in a natural way. 
In section 4, we consider the generalization to $q$-$W_n$ algebra.

\section{Root of Unity Limit of $q$-Virasoro Algebra}
In this section, we review the root of unity limit \cite{IOY2} of 
 the $q$-deformed Virasoro algebra \cite{SKAO} which has two parameters 
 $q$ and $t=q^{\beta}$. 
The defining relation is  
\be
f(z'/z) \mathcal{T}(z) \mathcal{T}(z') 
- f(z/z') \mathcal{T}(z') \mathcal{T}(z)
= \frac{(1-q)(1-t^{-1})}{(1-p)}
\Bigl[ \delta(pz/z') - \delta(p^{-1}z/z') \Bigr],
\ee
where $p = q/t$ and 
\be
f(z) = \exp\left( \sum_{n=1}^{\infty} \frac{1}{n}
\frac{(1-q^n)(1-t^{-n})}{(1+p^n)} z^n \right). 
\ee
The multiplicative delta function is defined by
\be
\delta(z) = \sum_{n \in \mathbb{Z}} z^n.
\ee
Using the $q$-deformed Heisenberg algebra $\mathcal{H}_{q,t}$:
\be
\begin{split}
[ \alpha_n, \alpha_m ] &= - \frac{1}{n} 
\frac{(1-q^n)(1-t^{-n})}{(1+p^n)} \delta_{n+m,0}, \qq (n \neq 0), \cr
[ \alpha_n, Q ] &= \delta_{n,0},
\end{split}
\ee
the $q$-Virasoro operator ${\mathcal T}(z)$ can be realized as 
\bel{qVg}
\mathcal{T}(z) = : \exp\left( \sum_{ n \neq 0} \alpha_n z^{-n} \right):
p^{1/2} q^{\sqrt{\beta} \alpha_0}
+ : \exp\left( - \sum_{n \neq 0} \alpha_n (pz)^{-n} \right):
p^{-1/2} q^{-\sqrt{\beta} \alpha_0},
\ee
The $q$-deformed chiral bosons are defined 
in terms of the $q$-deformed Heisenberg algebra as 
\be
\widetilde{\varphi}^{(\pm)}(z) = \widetilde{\varphi}^{(\pm)}_0(z)
+ \widetilde{\varphi}^{(\pm)}_R(z),
\ee
where
\be
\begin{split}
\widetilde{\varphi}^{(\pm)}_0(z) &= \beta^{\pm 1/2} Q
+ \frac{2}{r} \beta^{\pm 1/2} \alpha_0 \log z^r
+ \sum_{n \neq 0} \frac{(1+p^{-nr})}{(1-\xi_{\pm}^{nr})} \alpha_{nr} \, z^{-nr},
\cr
\widetilde{\varphi}^{(\pm)}_R(z)
&= \sum_{\ell=1}^{r-1} \sum_{n \in \mathbb{Z}}
\frac{(1+p^{-nr-\ell})}{1-\xi_{\pm}^{nr+\ell}} \alpha_{nr+\ell} \, z^{-nr-\ell}.
\end{split}
\ee
Here $\xi_+ = q$, $\xi_- = t$.

Let us consider the simultaneous $r$-th root of unity limit in $q$ and $t$
 which is given by   
\be
 q = \omega \ex^{-\frac{1}{\sqrt{\beta}} h}, ~~~~~
 t = \omega \ex^{-\sqrt{\beta} h}, ~~~~~
 p = \ex^{Q_E h}, ~~~~~
 h \to 0, 
\ee 
where $\omega = \ex^{\frac{2\pi \im}{r}}$ and 
 $Q_E = \sqrt{\beta} - \frac{1}{\sqrt{\beta}}$. 
Since $t = q^{\beta}$, 
this limit is possible if the parameter $\beta$ takes the rational number such as 
\be
 \beta = \frac{r m_- +1}{r m_+ +1}, 
 \label{beta:rational}
\ee
where $m_{\pm}$ are non-negative integers. 
In the limit, we have two types of bosons $\phi(w)$ and $\varphi(w)$ \cite{IOY2}
respectively given by
\be
\begin{split}
\lim_{h \rightarrow 0} \widetilde{\varphi}_0^{(\pm)}(z) &= \sqrt{\frac{2}{r}}\, 
\beta^{\pm 1/2} \phi(w), \cr
\lim_{h \rightarrow 0} \widetilde{\varphi}^{(\pm)}_R(z)
&= \sqrt{\frac{2}{r}} \, \varphi(w),
\end{split}
\ee
where $w=z^r$ and 
\be
\phi(w) = Q_0 + a_0 \log w  - \sum_{n \neq 0} \frac{a_n}{n} w^{-n},
\ee
\be
\varphi(w) = \sum_{\ell=1}^{r-1} \varphi^{(\ell)}(w),
\qq
\varphi^{(\ell)}(w) = \sum_{n \in \mathbb{Z}}
\frac{\tilde{a}_{n+\ell/r}}{n+\ell/r} w^{-n-\ell/r}.
\ee 
The commutation relations are 
\be
\begin{split}
 [a_m , a_n] = m \delta_{m+n,0}, ~~~~ [a_n,Q_0] = \delta_{n,0}, \\
 [\widetilde{a}_{n+\ell/r}, \widetilde{a}_{-m-\ell'/r}] 
 = (n+\ell/r) \delta_{m,m'} \delta_{\ell,\ell'}.
\end{split}
\ee
The boson $\phi(w)$ and the twisted boson $\varphi(w)$ play an important role 
 for the appearance of the ${\bf Z}_r$-parafermions.

\section{$\bm{Z_r}$-parafermionic CFT}
The $q$-deformed screening current and the charge are defined respectively by 
\be
 S^{(\pm)}(z) = :\ex^{\tilde{\varphi}^{(\pm)}(z)}:, ~~~~~~
 Q^{(\pm)}_{[a,b]} = \int_a^b \de_{\xi_{\pm}} z S^{(\pm)}(z), 
\ee  
where the Jackson integral is defined by 
\be
 \int_0^a \de \xi_{\pm}z f(z) = a(1-q)\sum_{k=0}^{\infty} f(aq^k)q^k.  
\ee
Multiplying the regularization factor, 
 we obtain the screening charge in the root of unity limit, 
 up to normalization, 
\be
 Q^{(\pm)}_{[a^r,b^r]}
 \equiv \lim_{h \to 0} \frac{(1-q^r)}{(1-q)} Q^{(\pm)}_{[a,b]}
 = \int_{a^r}^{b^r} \de w {\psi_1(w)} : e^{\sqrt{\beta} \phi(w)},
\ee
where we have defined \cite{CMM}
\be
 \psi_1(w) = \frac{A_r}{w^{(r-1)/r}} \sum_{k=0}^{r-1} \omega^{k} 
 :\exp \left\{ \sqrt{\frac{2}{r}} \phi^{(k)} (w) \right\}:.
 \label{para:1st}
\ee
Here $A_r$ is the normalization factor and 
we have introduced
\be
 \phi^{(k)} (w) \equiv \varphi( e^{2 \pi \im k} w). 
\ee
The correlation function is given by 
\be
 \langle \phi^{(k)} (w) \phi^{(k')} (w') \rangle = 
 \log 
 \frac{(1 - \omega^{k'-k} (w'/w)^{1/r})^r}{1-w'/w}
 =
 \log 
 \frac{(1-w'/w)^{r-1}}{\prod_{j=1}^{r-1} 
 (1 - \omega^{k'-k + j} (w'/w)^{1/r})^r}. 
\ee
Note that
\be
 \phi^{(k+1)} (w) =  \phi^{(k)} ( e^{2 \pi \im} w), ~~~~~~~
 \phi^{(r + k)}(w) = \phi^{(k)}(w), ~~~~~~~
 \sum_{k=0}^{r-1} \phi^{(k)} (w)  = 0.  
\ee

For example, we consider the $r=2$ case. In the limit, we obtain
\be
 \lim_{q \rightarrow -1} S(z)
= : \ex^{\sqrt{\beta} \phi(w)} \ex^{\varphi(w)}:, 
\ee
and after the appropriate normalization, 
 we obtain the following screening charge for the superconformal block 
 \cite{KIKKMO,AGZ}: 
\bel{screening charge}
 {Q_{[a^2,b^2]} = 
\int_{a^2}^{b^2} \de w \, {\psi(w)} : \ex^{\sqrt{\beta} \phi(w)} :}, 
\ee
where  
\be
 \psi(w)\equiv \frac{\im}{2 \sqrt{2 w}}
 \Bigl( : \ex^{\varphi(w)}: - : \ex^{-\varphi(w)} : \Bigr), ~~~~~
 \langle \psi(w_1) \psi(w_2) \rangle
 = \frac{1}{w_1-w_2}, 
\ee
is the NS fermion. 

 From now on we will show that the $\bm{Z}_r$-parafermions appear
 in the general $r$-th root of unity limit. 
In particular, $\psi_1(w)$ will be shown to work as the first parafermion current. 

The $\bm{Z}_r$-parafermion algebra consists of $(r-1)$ currents 
 $\psi_{\ell}(w)~ (\ell=1, \cdots, r-1)$ satisfying 
 the following defining relations \cite{FZ}: 
\begin{align}
 &\psi_{\ell}(w) \psi_{\ell'}(w') = \frac{c_{\ell,\ell'}}{(w-w')^{2\ell\ell'/r}} 
 \left\{ \psi_{\ell+\ell'}(w') + \mathcal{O}(w-w') \right\},~~~ \ell + \ell' < r, 
 \label{para1}\\
 &\psi_{\ell}(w) \psi^{\dag}_{\ell'}(w') = c_{\ell, r-\ell'} (w-w')^{-2\ell(r-\ell')/r}
 \left\{
  \psi_{\ell-\ell'}(w') + \mathcal{O}(w-w') 
 \right\}, ~~~ \ell' < \ell 
 \label{para2}\\
 &\psi_{\ell}(w) \psi_{\ell}^{\dag}(w') = (w-w')^{-2 \Delta_{\ell}} 
 \left\{
  1 + \frac{2 \Delta_{\ell} }{c_p} (w-w')^2 T_{\rm PF}(w) + \mathcal{O}((w-w')^3) 
 \right\}, 
 \label{para3}
\end{align}
where $\psi^{\dag}_{\ell}(w) = \psi_{r-\ell}(w)$ and 
\be
 \Delta_{\ell} = \frac{\ell(r-\ell)}{r}, ~~~~~
 c_p = \frac{2(r-1)}{r+2}, 
\ee
are the conformal dimension of $ \psi_{\ell}(w)$ and the central charge 
 of the parafermionic stress tensor $T_{\rm PF}$. 
The explicit form of $T_{\rm PF}(w)$ is given in \cite{Mar}. 
The coefficients $c_{\ell \ell'}$ are given by 
\be
 c_{\ell \ell'}
 = \sqrt{\frac{(\ell+\ell')! (r-\ell) ! (r-\ell')!}
 {\ell! \ell'! (r-\ell-\ell')! r!}}. 
\ee

The OPE of (\ref{para:1st}) is   
\begin{align}
 &\psi_1 (w) \psi_1 (w') 
 \equiv \frac{c_{1,1}}{(w-w')^{2/r}} \left\{ 
 \psi_2(w) + \mathcal{O}(w-w') 
 \right\}. 
\end{align}
Here we have defined the second parafermion,   
\be
 \psi_2(w) = \frac{A_r^2}{c_{1,1} w^{2(r-2)/r}} \sum_{k,k'=0}^{r-1} 
 \omega^{k+k'} (1 - \omega^{k'-k})^2  :\ex^{ \sqrt{\frac{2}{r}} 
 \left( \phi^{(k)}(w) + \phi^{(k')}(w) \right)}: 
\ee
Similarly, the $(\ell + 1)$-th parafermion is 
 obtained from $\ell$-th parafermion by  
\be
 \psi_{\ell+1}(w) \equiv \lim_{w' \to w} 
 \frac{(w - w')^{2\ell/r}}{ c_{1,\ell}} 
 \psi_1 (w') \psi_{\ell} (w). 
\ee
In particular, 
\be
 \psi^{\dag}_1 (w) \equiv \psi_{r-1}(w) 
 = \frac{B_r}{w^{(r-1)/r}}\sum_{\ell=0}^{r-1} \omega^{\ell} 
 \exp\left\{ - \sqrt{\frac{2}{r}} \phi^{(\ell)}(w) \right\}, 
\ee
where $B_r$ is a constant which can be determined by the relation 
\be
 \langle \psi_{1}(w) \psi_1^{\dag}(w') \rangle 
 = \frac{1}{(w-w')^{2(r-1)/r}}. 
\ee 
After all, we have the chiral boson $\phi(w)$ coupled to $Q_E$ 
 and the $\bm{Z}_r$-parafermion $\psi_{\ell}(w)$. 
Therefore, the stress tensor of the whole system is 
\be
 T(w) = T_{\rm{B}}(w) + T_{\rm{PF}}(w), 
\ee
where $T_B(w)$ stands for the usual stress tensor for the chiral boson field. 
The central charge is 
\be
 c^{(r)} = 1 - \frac{6Q_E^2}{r} + \frac{2(r-1)}{r+2} 
         = \frac{3r}{r+2} - \frac{6 Q_E^2}{r}. 
 \label{cc:n=2}
\ee
Because $\beta$ is restricted to the rational number (\ref{beta:rational}), 
(\ref{cc:n=2}) is written as 
\be
 c^{(r,m,s)} = \frac{3r}{r+2} - \frac{6rs^2}{m(m+rs)}.
\ee 
where we have set $m = r m_+ +1$ and $s = m_- - m_+$. 
Especially, when $s = 1$, 
\be
 c^{(r,m,1)} = \frac{3r}{r+2} - \frac{6r}{m(m+r)}, 
\ee 
 is the  central charge of 
 the unitary series of the ${\bf Z}_r$-parafermionic CFT \cite{Zam}. 

The form of the screening charge in the case of general $r$ is the same as 
 that of eq. (\ref{screening charge}).

\section{Root of Unity Limit of $\bm{q}$-$\bm{W_n}$ Algebra}
In this section, we consider the generalization to the $q$-$W_n$ algebra
\cite{AKOS95}. 
We denote by $\mathfrak{h}$ 
 the Cartan subalgebra of $\mathfrak{sl}(n)$ Lie algebra. 
The $q$-$W_n$ algebra is expressed  
 in terms of the following $\mathfrak{h}$-valued $q$-deformed boson,
\be
 \langle e_a, {\widetilde{\varphi}^{(\pm)}(z)} \rangle \equiv 
 \widetilde{\varphi}_{a}^{(\pm)}(z)
 = \widetilde{\varphi}_{0,a}^{(\pm)}(z) + \widetilde{\varphi}_{R,a}^{(\pm)}(z), 
\ee
where
\begin{align}
  \widetilde{\varphi}^{(\pm)}_{0,a}(z) &= 
 \beta^{\pm\frac{1}{2}} Q_a 
 + \beta^{\pm\frac{1}{2}} \alpha_{0,a} \log z 
 + \sum_{n \neq 0} \frac{1}{\xi_{\pm}^{nr/2} - \xi_{\pm}^{-nr/2}} 
 \alpha_{nr,a} z^{-nr}, \\
 \widetilde{\varphi}_{R,a}^{(\pm)} (z) &= 
 \sum_{\ell=1}^{r-1} \widetilde{\varphi}_{\ell,a}^{(\pm)} (z) 
 = \sum_{\ell=1}^{r-1} \sum_{n \in {\bf Z}} 
 \frac{1}{\xi_{\pm}^{(nr+\ell)/2} - \xi_{\pm}^{-(nr+\ell)/2}} 
 \alpha_{nr+\ell,a} z^{-(nr+\ell)}, 
\end{align}
and $e_a~ (a=1, \cdots, n-1)$ are the simple roots 
 and $\langle , \rangle: {\mathfrak h}^* \otimes {\mathfrak h} \to {\bf C}$
 is the canonical pairing. 
The commutation relations are given by 
\be
\begin{split}
&[Q_a,\alpha_{0,b}]=C_{ab}, \\
 &[\alpha_{n,a},\alpha_{m,b}] = \frac{1}{n}(q^{n/2}-q^{-n/2})(t^{n/2}-t^{-n/2})
 C_{ab}(p) \delta_{n+m,0}, \\
 &[Q_a,Q_b]=0, ~~~~[\alpha_{0,a},\alpha_{0,b}]=0, 
\end{split}
\ee
where $C_{ab}$ is the Cartan matrix of $A$ type and
\be
 C_{ab}(p) = [2]_{p} \delta_{a,b} -p^{1/2}\delta_{a,b-1} -p^{-1/2}\delta_{a-1,b}. 
\ee
The $q$-number is defined by 
\be
 [n]_q = \frac{q^{n/2}-q^{-n/2}}{q^{1/2}-q^{-1/2}}. 
\ee 
Similar to the $q$-Virasoro case, we consider the limit, 
\be
 q = \omega^k \ex^{-\frac{h}{\sqrt{\beta}}}, ~~~~
 t = \omega^k \ex^{-\sqrt{\beta}h}, ~~~~
 p=q/t = \ex^{Q_E h}, ~~~~ \omega=\ex^{\frac{2 \pi \im}{r}}, ~~~~~
 h \to +0,  
 \label{limit:general}
\ee
where $\omega = \ex^{\frac{2\pi \im}{r}}$ and 
 $k$ is a natural number mutually prime to $r$. 
The condition to be able to take this limit is that $\beta$ is a rational number,  
\be
 \beta = \frac{r m_- + k}{r m_+ + k}, 
 \label{beta:general}
\ee
where $m_{\pm}$ are non-negative integers. 
Taking this limit, 
\begin{align}
 &\lim_{h \to 0} \widetilde{\varphi}_0^{a}(z) 
 = \frac{1}{\sqrt{r}}\beta^{1/2} \phi^a(w), \\
 &\lim_{h \to 0} \widetilde{\varphi}_R^{a}(z) 
 = \frac{1}{\sqrt{r}} \varphi^a(w), 
\end{align}
we obtain 
\begin{align}
 &\phi^a(w) = Q_0^a + a_0^a \log w - \sum_{n \neq 0} \frac{1}{n} a_n^a w^{-n}, \\
 &\varphi^a(w) = \sum_{\ell = 1}^{r-1} \varphi_{\ell}(w), ~~~~~
 \varphi_{\ell}(w) = 
 \sum_{\ell = 1}^{r-1} \sum_{n \in {\bf Z}} \frac{1}{n+\ell/r} 
 \widetilde{a}_{n+\ell/r}^a w^{-(n+\ell/r)}, 
\end{align}
Here we have normalized as 
\begin{align}
 &Q^a = \frac{1}{\sqrt{r}} Q^a_0, ~~~~
 \alpha_0^a = \sqrt{r} a_0^a , \\
 &\alpha_{nr}^a = - (-1)^{nk} \sqrt{r} h a_n^a, \\
 &\alpha_{nr+\ell}^a 
 = \frac{e^{\im \pi k(nr+\ell)/2}-e^{-\im \pi k(nr+\ell)/2}}{\sqrt{r}(n+\ell/r)} 
 \widetilde{a}_{n+\ell/r}^a. 
\end{align}
The commutation relations are 
\begin{align}
 &[Q^a,\alpha_0^b]=C_{ab}, ~~~~[Q^a,Q^b]=0, ~~~~[\alpha_0^a,\alpha_0^b]=0, \\
 &[a_n^a,a_m^b] = n C_{ab} \delta_{n+m,0}, \\
 &[\widetilde{a}_{n+\ell/r}^a,\widetilde{a}_{-m-\ell'/r}^b] 
 = \left( n + \frac{\ell}{r} \right) C_{ab} \delta_{n,m} \delta_{\ell,\ell'}. 
\end{align} 
The correlation functions are 
\begin{align}
 &\langle \phi^a(w) \phi^b(w') \rangle = C_{ab} \log (w-w'), \\
 &\langle \varphi_{\ell}^a(w) \varphi_{\ell'}^b(w') \rangle 
 = \delta_{\ell+\ell',r} C_{ab} \sum_{k=0}^{r-1} \omega^{-k\ell}
  \log \left[ 1 - \omega^k \left( \frac{w'}{w} \right)^{\frac{1}{r}} \right], \\
 &\langle \varphi^a(w) \varphi^b(w') \rangle 
 = C_{ab}
  \log \left[ \frac{(1 - (w'/w)^{1/r})^r}{1 - (w'/w)} \right]. 
\end{align}
For each $e_a$, we define
\be
 \psi_{e_a} (w) = \frac{A_r}{w^{(r-1)/r}} \sum_{\ell=0}^{r-1} \omega^{\ell} 
 :\exp \left[ \sqrt{\frac{1}{r}} \phi^{(\ell)}_a (w) \right]:, 
 \label{parafermion:first}
\ee
where $A_r$ is a normalization factor and 
\be
 \phi_a^{(\ell)} (w) \equiv \varphi_a( e^{2 \pi \im \ell} w). 
\ee
Let $\alpha = \sum_{a=1}^{n-1} n_a e_a \in Q$, 
 where $n_a$ are non-negative integers and 
 $Q$ denotes the root lattice. 
We obtain the corresponding parafermion, up to its normalization, 
\begin{align}
 \psi_{\alpha} \sim \prod \psi_{e_a}^{n_a}. 
\end{align}
The independent parafermion can be given only for the case 
$\alpha \in Q/r Q$.  
Not of all $\psi_{\alpha}$ are independent; 
\be
 1 \sim \underbrace{ \psi_{e_a} \cdots \cdots \psi_{e_a}}_{r}. 
\ee
For example, in the the case of $\mathfrak{sl}(3)$ algebra and $r=4$,  
 the corresponding parafermions are drawn in the Fig. 1. 
We define the parafermion associated with negative of a simple root by 
\begin{align}
 \psi_{-e_a} \sim \underbrace{\psi_{e_a} \psi_{e_a} \cdots \psi_{e_a}}_{r-1}.  
\end{align}
The normalization can be determined by 
the correlation functions \cite{Gep},
\begin{align}
 &\langle \psi_{\alpha}(w) \psi_{-\alpha}(w') \rangle = 
 (w-w')^{-2 + \frac{\alpha^2}{r}}, 
\end{align} 
where $\alpha^2=(\alpha,\alpha)$. 
In particular, 
\be
 \langle \psi_{e_a}(w) \psi_{-e_a}(w') \rangle = 
 (w-w')^{-2\frac{r-1}{r}}. 
\ee
In the case of the $\mathfrak{sl}(2)$ algebra, we obtain the first 
 ${\bf Z}_r$-parafermion, 
\be
 \psi_1(w) = \psi_{e_1}(w).
\ee

\begin{figure}[h]
\begin{center}
 \includegraphics[height=6cm]{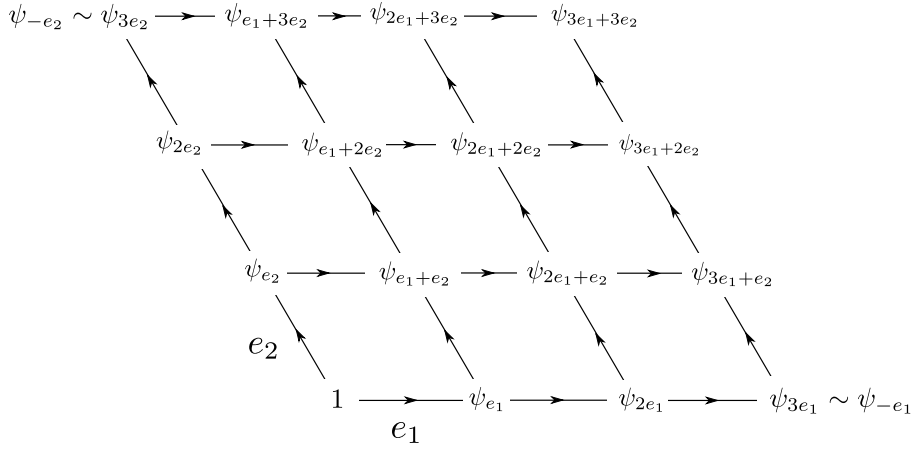}
\caption{The parafermions in the case of $\mathfrak{sl}(3)$ and $r=4$.}
\end{center}
\end{figure}

Similar to the case of $n=2$ (\ref{cc:n=2}), the central charge is given by 
\begin{align}
 c_n^{(r)} &= \frac{n(n-1)(r-1)}{r+n} 
 + (n-1)\left(1 - n(n+1) \frac{Q_E^2}{r}\right) \notag\\
 &= \frac{r(n^2-1)}{r+n}  - n(n^2-1)\frac{Q_E^2}{r}. 
\end{align}
When we set $m = r m_+ + k$, $m_- = m_+ + s$ in (\ref{beta:general}), 
 this central charge becomes 
\begin{align}
 c^{(r,m,s)}_n &= \frac{r (n^2-1)}{r + n}
- \frac{rs^2n(n^2-1)}{m(m+rs)} \cr
&= \frac{(n^2-1)r(\frac{m}{s}-n)(\frac{m}{s}+n+r)}
{(r+n)\frac{m}{s}(\frac{m}{s}+r)}, 
\label{cc:rms}
\end{align}
which is the same as that of the coset model, 
\be
 \frac{\widehat{\mathfrak{sl}}(n)_{r} \oplus 
 \widehat{\mathfrak{sl}}(n)_{\frac{m}{s}-n}}
 {\widehat{\mathfrak{sl}}(n)_{\frac{m}{s}-n+r}}. 
\ee
Compared with (\ref{coset}) we find 
\be
 p = \frac{m}{s} - n. 
\ee

In the case of $s=0$ corresponding to $Q_E=0$, 
we have the central charge of the usual Sugawara stress tensor for 
 $\widehat{\mathfrak{\mathfrak{sl}}}(n)_{r}$,  
\be
 c_n^{(r,m,0)} = \frac{r (n^2-1)}{r+n} 
 = c_{\widehat{\mathfrak{sl}}(n)_{r}}
\ee  
It is well-known that 
the affine Lie algebra $\widehat{\mathfrak{sl}}(n)_r$ is represented 
by parafermions and an auxiliary boson \cite{Gep}. 
In the case of $s=1$, 
because (\ref{cc:rms}) becomes
\be 
 c^{(r,m,1)}_n = \frac{(n^2-1)r(m-n)(m+n+r)}{(r+n)m(m+r)}, 
\ee 
the model gives us the unitary series of the coset,  
\be
 \frac{\widehat{\mathfrak{sl}}(n)_r \oplus \widehat{\mathfrak{sl}}(n)_{m-n}}
 {\widehat{\mathfrak{sl}}(n)_{m-n+r}}. 
\ee

We can see how the level $p$ 
 is related with the omega-background parameters $\epsilon_{1}$ and $\epsilon_2$
 in the $4$-d side. 
Since $\beta = - \epsilon_1/\epsilon_2$, 
 (\ref{beta:general}) yields the condition to the ratio of these parameters. 
Therefore, when we introduce the free parameter $\epsilon$, 
 $\epsilon_{1,2}$ can be written respectively as 
\be
 \epsilon_1 = \epsilon (p + n + r), ~~~~
 \epsilon_2 = -\epsilon (p + n).
\ee
This result suggests that 
 the Nekrasov-Shatashvili limit 
 $\epsilon_1 \to 0$ (resp. $\epsilon_2 \to 0$) of 
 the $\mathcal{N}=2$ gauge theory on the ${\bf R}^4/{\bf Z}_r$ corresponds to 
 the critical level limit $p + r \to -n$ (resp.  $p \to -n$) of 
 the coset model.

\section*{Acknowledgments}
We thank D. Serban for valuable discussions. 
The authors' research is supported in part by the
Grant-in-Aid for Scientific Research from the Ministry of Education, 
Science and Culture, Japan(23540316).




\end{document}